\documentclass[journal]{IEEEtran}

\usepackage{amsmath,amsfonts}
\usepackage{algorithmic}
\usepackage{algorithm}
\usepackage{array}
\usepackage[caption=false,font=normalsize,labelfont=sf,textfont=sf]{subfig}
\usepackage{textcomp}
\usepackage{stfloats}
\usepackage{url}
\usepackage{verbatim}
\usepackage{graphicx}
\usepackage{cite}

\usepackage{etoolbox}
\makeatletter
\apptocmd\@makecaption{\par}{}{%
  \errmessage{\noexpand\@makecaption could not be patched}%
}
\makeatother

\usepackage[final]{listings}
\usepackage{xcolor}
\usepackage{pythonhighlight}
\usepackage{siunitx}
\usepackage{makecell}

\begin{document}

% format SI ranges more concisely
\sisetup{range-phrase=--}
\sisetup{range-units=single}

% ref: https://tex.stackexchange.com/questions/377122/typesetting-for-a-verilog-lstinput

\definecolor{verilogcommentcolor}{RGB}{104,180,104}
\definecolor{verilogkeywordcolor}{RGB}{49,49,255}
\definecolor{verilogsystemcolor}{RGB}{128,0,255}
\definecolor{verilognumbercolor}{RGB}{255,143,102}
\definecolor{verilogstringcolor}{RGB}{160,160,160}
\definecolor{verilogdefinecolor}{RGB}{128,64,0}
\definecolor{verilogoperatorcolor}{RGB}{0,0,128}

% Verilog style
\lstdefinestyle{prettyverilog}{
   language           = Verilog,
   commentstyle       = \color{verilogcommentcolor},
   alsoletter         = \$'0123456789\`,
   literate           = *{+}{{\verilogColorOperator{+}}}{1}%
                         {-}{{\verilogColorOperator{-}}}{1}%
                         {@}{{\verilogColorOperator{@}}}{1}%
                         {;}{{\verilogColorOperator{;}}}{1}%
                         {*}{{\verilogColorOperator{*}}}{1}%
                         {?}{{\verilogColorOperator{?}}}{1}%
                         {:}{{\verilogColorOperator{:}}}{1}%
                         {<}{{\verilogColorOperator{<}}}{1}%
                         {>}{{\verilogColorOperator{>}}}{1}%
                         {=}{{\verilogColorOperator{=}}}{1}%
                         {!}{{\verilogColorOperator{!}}}{1}%
                         {^}{{\verilogColorOperator{^}}}{1}%
                         {|}{{\verilogColorOperator{|}}}{1}%
                         {=}{{\verilogColorOperator{=}}}{1}%
                         {[}{{\verilogColorOperator{[}}}{1}%
                         {]}{{\verilogColorOperator{]}}}{1}%
                         {(}{{\verilogColorOperator{(}}}{1}%
                         {)}{{\verilogColorOperator{)}}}{1}%
                         {,}{{\verilogColorOperator{,}}}{1}%
                         {.}{{\verilogColorOperator{.}}}{1}%
                         {~}{{\verilogColorOperator{$\sim$}}}{1}%
                         {\%}{{\verilogColorOperator{\%}}}{1}%
                         {\&}{{\verilogColorOperator{\&}}}{1}%
                         {\#}{{\verilogColorOperator{\#}}}{1}%
                         {\ /\ }{{\verilogColorOperator{\ /\ }}}{3}%
                        ,
   morestring         = [s][\color{verilogstringcolor}]{"}{"},%
   identifierstyle    = \color{black},
   vlogdefinestyle    = \color{verilogdefinecolor},
   vlogconstantstyle  = \color{verilognumbercolor},
   vlogsystemstyle    = \color{verilogsystemcolor},
   basicstyle         = \scriptsize\fontencoding{T1}\ttfamily,
   keywordstyle       = \bfseries\color{verilogkeywordcolor},
   numbers            = left,
   numbersep          = 10pt,
   tabsize            = 4,
   escapechar         = £,
   upquote            = true,
   sensitive          = true,
   showstringspaces   = false, %without this there will be a symbol in the places where there is a space
   frame              = single
}

% This is shamelessly stolen and modified from:
% https://github.com/jubobs/sclang-prettifier/blob/master/sclang-prettifier.dtx
\makeatletter

% Language name
\newcommand\language@verilog{Verilog}
\expandafter\lst@NormedDef\expandafter\languageNormedDefd@verilog%
  \expandafter{\language@verilog}
  
% save definition of single quote for testing
\lst@SaveOutputDef{`'}\quotesngl@verilog
\lst@SaveOutputDef{``}\backtick@verilog
\lst@SaveOutputDef{`\$}\dollar@verilog

% Extract first character token in sequence and store in macro 
% firstchar@verilog, per http://tex.stackexchange.com/a/159267/21891
\newcommand\getfirstchar@verilog{}
\newcommand\getfirstchar@@verilog{}
\newcommand\firstchar@verilog{}
\def\getfirstchar@verilog#1{\getfirstchar@@verilog#1\relax}
\def\getfirstchar@@verilog#1#2\relax{\def\firstchar@verilog{#1}}

% Initially empty hook for lst
\newcommand\addedToOutput@verilog{}
\lst@AddToHook{Output}{\addedToOutput@verilog}

% The style used for constants as set in lstdefinestyle
\newcommand\constantstyle@verilog{}
\lst@Key{vlogconstantstyle}\relax%
   {\def\constantstyle@verilog{#1}}

% The style used for defines as set in lstdefinestyle
\newcommand\definestyle@verilog{}
\lst@Key{vlogdefinestyle}\relax%
   {\def\definestyle@verilog{#1}}

% The style used for defines as set in lstdefinestyle
\newcommand\systemstyle@verilog{}
\lst@Key{vlogsystemstyle}\relax%
   {\def\systemstyle@verilog{#1}}

% Counter used to check current character is a digit
\newcount\currentchar@verilog
  
% Processing macro
\newcommand\@ddedToOutput@verilog
{%
   % If we're in \lstpkg{}' processing mode...
   \ifnum\lst@mode=\lst@Pmode%
      % Save the first token in the current identifier to \@getfirstchar
      \expandafter\getfirstchar@verilog\expandafter{\the\lst@token}%
      % Check if the token is a backtick
      \expandafter\ifx\firstchar@verilog\backtick@verilog
         % If so, then this starts a define
         \let\lst@thestyle\definestyle@verilog%
      \else
         % Check if the token is a dollar
         \expandafter\ifx\firstchar@verilog\dollar@verilog
            % If so, then this starts a system command
            \let\lst@thestyle\systemstyle@verilog%
         \else
            % Check if the token starts with a single quote
            \expandafter\ifx\firstchar@verilog\quotesngl@verilog
               % If so, then this starts a constant without length
               \let\lst@thestyle\constantstyle@verilog%
            \else
               \currentchar@verilog=48
               \loop
                  \expandafter\ifnum%
                  \expandafter`\firstchar@verilog=\currentchar@verilog%
                     \let\lst@thestyle\constantstyle@verilog%
                     \let\iterate\relax%
                  \fi
                  \advance\currentchar@verilog by \@ne%
                  \unless\ifnum\currentchar@verilog>57%
               \repeat%
            \fi
         \fi
      \fi
      % ...but override by keyword style if a keyword is detected!
      %\lsthk@DetectKeywords% 
   \fi
}

% Add processing macro only if verilog
\lst@AddToHook{PreInit}{%
  \ifx\lst@language\languageNormedDefd@verilog%
    \let\addedToOutput@verilog\@ddedToOutput@verilog%
  \fi
}

% Colour operators in literate
\newcommand{\verilogColorOperator}[1]
{%
  \ifnum\lst@mode=\lst@Pmode\relax%
   {\bfseries\textcolor{verilogoperatorcolor}{#1}}%
  \else
    #1%
  \fi
}

\makeatother

\title{Switchboard: An Open-Source Framework for Modular Simulation of Large Hardware Systems}

\author{Steven Herbst, Noah Moroze, Edgar Iglesias, Andreas Olofsson
\thanks{S. Herbst and A. Olofsson are with Zero ASIC Corporation, Lexington, MA 02421 (email: andreas@zeroasic.com).  N. Moroze and E. Iglesias were with the same company when they contributed to this work.}\thanks{This work has been submitted to the IEEE for possible publication. Copyright may be transferred without notice, after which this version may no longer be accessible.}}

% \markboth{IEEE Transactions on Computer-Aided Design of Integrated Circuits and Systems,~Vol.~??, No.~??,~Month~Year}{Herbst \MakeLowercase{\textit{et al.}}: Switchboard}

\maketitle

\begin{abstract}
Scaling up hardware systems has become an important tactic for improving performance as Moore's law fades.  Unfortunately, simulations of large hardware systems are often a design bottleneck due to slow throughput and long build times.  In this article, we propose a solution targeting designs composed of modular blocks connected by latency-insensitive interfaces.  Our approach is to construct the hardware simulation in a similar fashion as the design itself, using a prebuilt simulator for each block and connecting the simulators via fast shared-memory queues at runtime.  This improves build time, because simulation scale-up simply involves running more instances of the prebuilt simulators.  It also addresses simulation speed, because prebuilt simulators can run in parallel, without fine-grained synchronization or global barriers. 
 We introduce a framework, Switchboard, that implements our approach, and discuss two applications, demonstrating its speed, scalability, and accuracy: (1) a web application where users can run fast simulations of chiplets on an interposer, and (2) a wafer-scale simulation of one million RISC-V cores distributed across thousands of cloud compute cores.
\end{abstract}


\section{Introduction}

\IEEEPARstart{S}{cale-up} is one of the key tactics for improving hardware performance in the post-Moore's law era.  It is applicable to system design at a broad range of hierarchical levels, for example by growing chip designs to the reticle limit, wafer-scale integration of chip designs, integration of chiplets on interposers, and PCB- and server rack-level integration.  Modern GPU-based accelerators already use many of these techniques.

From a design perspective, a challenge of scale-up is that simulating very large hardware systems is slow, both in terms of the time required to build simulations, and the runtime speed of simulations.  This is particularly problematic because large hardware systems are often running intricate algorithms that take vast numbers of clock cycles to complete, such as artificial neural networks.

Parallel simulation of register-transfer level (RTL) sources is one way to speed up large simulations, and there has been some exciting progress on this front in the last few years.  A common approach is to focus on full-cycle simulation, constructing a computational graph for an entire hardware system, and then partitioning the graph among compute cores on a host system.  This type of RTL simulation parallelizes well, sometimes yielding superlinear speedups as more cores are allocated to the task~\cite{repcut}.

However, there are some drawbacks to direct parallelization of a large RTL design.  As the size of a hardware system grows, the simulator build system has to read in and manipulate a larger and larger design, slowing down the build process.   In addition, parallel simulation threads generally need to synchronize on every cycle, requiring accurate load balancing to prevent threads from sitting idle; the load balancing process can in turn increase build time.  Finally, parallel RTL simulation approaches are generally geared towards running on a single host system with many cores, meaning that in practice simulations can only be parallelized on a few dozen to a few hundred cores \cite{repcut} or require special compute hardware that may have limited availability \cite{parendi}.

In this work, we take a different approach that is a combination of a design methodology and a simulation technique.  We focus on hardware systems composed of reasonably-sized blocks that communicate with each other through latency-insensitive interfaces, a time-tested design methodology that offers many benefits~\cite{latency-insensitive}.  We build a simulator for each unique block, which is fast because the blocks are reasonably-sized.  At runtime, a prebuilt simulator is launched for each block instance, and block simulations communicate through shared-memory queues that convey latency-insensitive interfaces.

Simulations don't have to be explicitly synchronized for functional correctness, due to the fact that their communication channels are latency-insensitive.  This makes it straightforward to distribute simulations across large numbers of cloud compute instances, breaking through the typical requirement for a single compute instance with many cores.  In doing this, we give up the expectation of cycle accuracy, but it is still possible to gather approximate performance measurements, as we describe later.

We developed a software framework, Switchboard, that makes it easy to implement our modular approach to large-scale hardware simulation.  Switchboard is a Python package that provides automation for building modular simulations and connecting them together at runtime using a high-performance shared-memory queue implementation.  It also provides tools for building software-based (non-RTL) models of hardware blocks for rapid prototyping.

This article is organized as follows: Section~\ref{sec:modular} details our approach to modular simulation, and Section~\ref{sec:switchboard} describes the Switchboard framework, showing examples of how it can be used to implement our approach.  In Section~\ref{sec:applications}, we evaluate Switchboard with two applications: (1) a web app where chiplets can be arranged on a silicon interposer and simulated interactively, and (2) a wafer-scale simulation of one million RISC-V cores using standard cloud compute resources.  Section~\ref{sec:related} describes related work, and Section~\ref{sec:conclusion} concludes.

\section{Modular Design and Simulation}
\label{sec:modular}

Switchboard is based on the philosophy that large hardware systems should be composed of repeated instances of reasonably-sized modular blocks that communicate through latency-insensitive channels (AXI~\cite{axi}, TileLink~\cite{tilelink}, UMI~\cite{umi}, etc.).  Here, the meaning of ``block'' is flexible: it could be part of a chip implementation, or it could be a chiplet sitting on an interposer. 
 We consider a block to be reasonably-sized if its corresponding simulator can be built quickly and runs with high performance for the context in which it is being used.

\begin{figure}
    \centering
    \includegraphics[width=\linewidth]{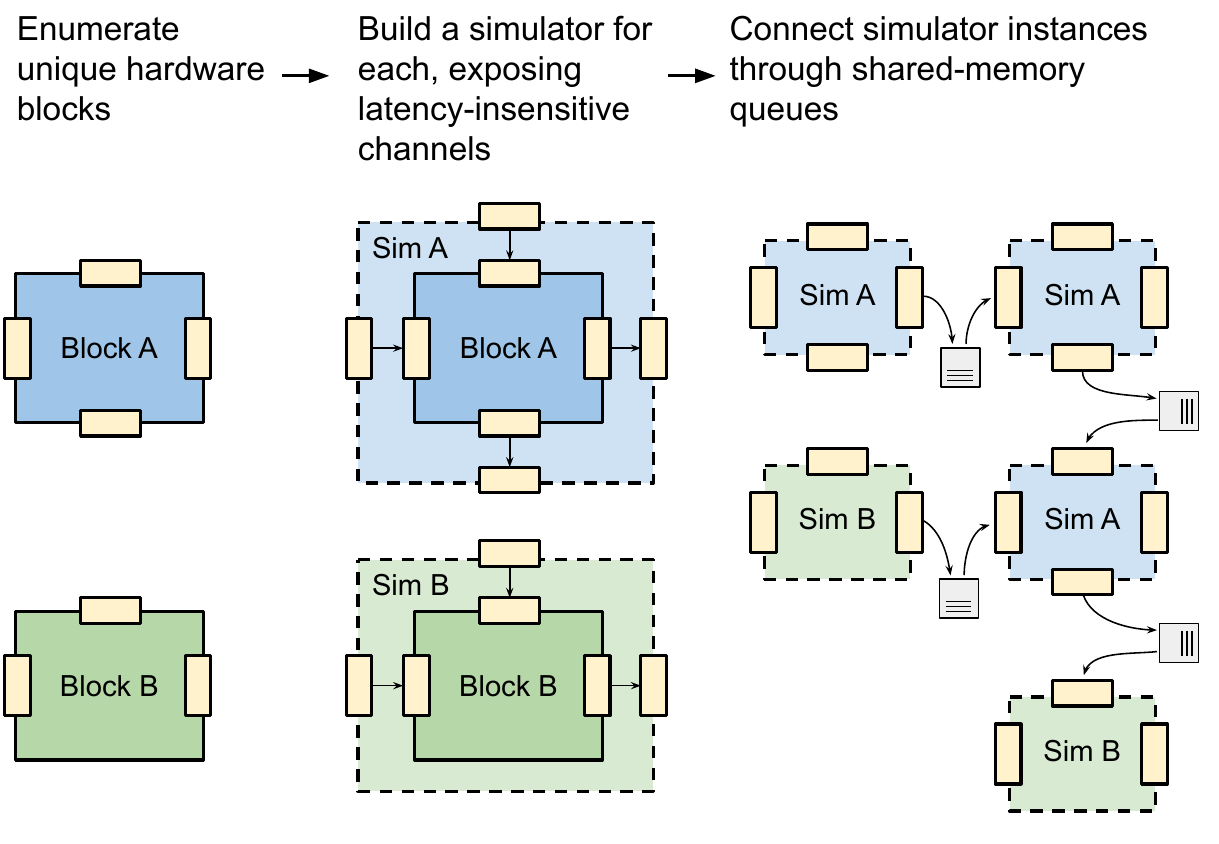}
    \caption{Our modular simulation approach.}
    \label{fig:main-idea}
\end{figure}

Our approach to simulating a design constructed this way is illustrated in Figure~\ref{fig:main-idea}.  We start by enumerating the unique modular blocks in a design.  Each unique block is wrapped by tying off its latency-insensitive channels to bridge modules that connect to single-producer, single-consumer (SPSC) shared-memory queues.  A simulator is built for each wrapped block, after which point full system simulations can be constructed by running multiple instances of each block simulator, with transmit/receive pairs of channels ``wired together'' by configuring the corresponding bridge modules to point to the same shared memory queues at runtime.

\subsection{Queue-Based Communication}

\begin{figure}
    \centering
    \includegraphics[width=\linewidth]{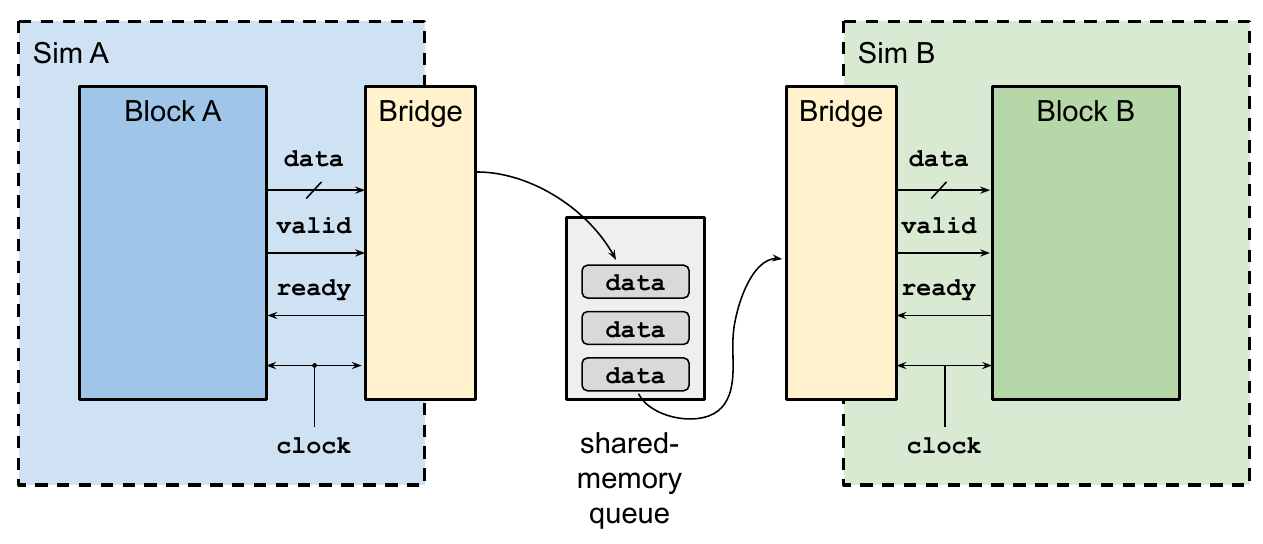}
    \caption{Conveying a latency-insensitive interface through a shared-memory queue.}
    \label{fig:queue-comms}
\end{figure}

A shared-memory queue is used to convey a latency-insensitive channel between two blocks, as illustrated in Figure~\ref{fig:queue-comms}.  Although other types of flow control could be used (e.g., credit-based flow), we assume for the purpose of discussion that the channel uses ready-valid handshaking, with a \textbf{data} bus and \textbf{valid} signal driven in one direction, and a \textbf{ready} signal sent in the opposite direction.  In this flow-control scheme, the transmitter asserts \textbf{valid} on every cycle that the data bus is valid, and the receiver asserts \textbf{ready} on every cycle that it can receive data.  If \textbf{valid} and \textbf{ready} are asserted in the same cycle, data is considered to have been transferred from the transmitter to the receiver; we refer to this as a ``handshake.''

On the transmitter side, the value of the \textbf{data} bus is stored in a packet and pushed into the queue when the bridge handshakes with the transmitter.  On the receiver side, the bridge removes a packet from the queue and drives the data bus with the value contained in the packet, with \textbf{valid} asserted.  When the receiver handshakes with the bridge, the bridge can move onto the next packet in the queue.

A queue is a natural choice for conveying a latency-insensitive channel because it propagates backpressure.  If the transmitter has nothing to send, eventually the queue will empty; the \textbf{valid} signal on the receiver side will then be de-asserted by the bridge.  Similarly, if the receiver is inactive, eventually the queue will fill and the \textbf{ready} signal on the transmitter side will be de-asserted by the bridge.

Shared-memory queues have two additional benefits.  First, they are fast, since system calls are only needed for the initial setup, not to move data while a simulation is running.  Second, they are straightforward to implement for a variety of hardware model implementations, as discussed next.

\subsection{Model Implementation}

RTL simulation is not the only option for modeling a hardware block.  For example, RTL code may be implemented on a field-programmable gate array (FPGA) to improve simulation speed.  If RTL code is not available, a hardware block may be modeled directly in software (e.g., with C++ or Python).  Analog modeling (e.g., SPICE) is also possible if analog-to-digital (A2D) and digital-to-analog (D2A) converters are used to connect analog pins to digital latency-insensitive interfaces.

\begin{figure}
    \centering
    \includegraphics[width=1\linewidth]{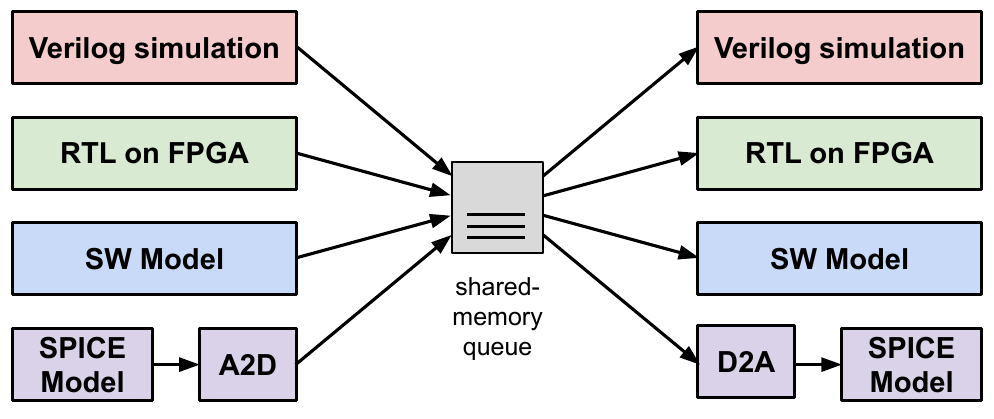}
    \caption{Standardizing communication around shared-memory queues makes it straightforward for any type of hardware model to communicate with any other type of hardware model.}
    \label{fig:interaction-matrix}
\end{figure}

As illustrated in Figure~\ref{fig:interaction-matrix}, our approach enables any type of model to communicate with any other type of model.  The key is that shared memory queues are the only mechanism by which models interact with each other.  This means that each model type only requires an adapter to connect it to a queue, rather than requiring an adapter for every pair of model types.  Hence, implementation work grows linearly instead of quadratically with the number of model types.

For FPGA emulation, we assume that one or more FPGAs are connected to a host system and are able to read and write sections of memory visible to the host.  An example of this type of system is an Amazon Web Services Elastic Compute Cloud (AWS EC2) F1 instance, which consists of 1-8 FPGAs connected to a host system via PCIe.

In such a setup, shared memory queues reside in host memory, and block models implemented on FPGAs interact with these queues over PCIe.  Models running on the host CPU can interact with queues without knowing that an FPGA is on the other side.  Alternatively, two FPGAs can communicate with each other through queues residing in host memory.

\subsection{Performance Simulation}
\label{sec:perfsim}

The modular simulation approach described yields functionally correct results as-is, but requires modification to estimate performance.  This is due to the fact that models run at different rates, and because there is real-world latency in sending data from one model to another.

\begin{figure}
    \centering
    \includegraphics[width=\linewidth]{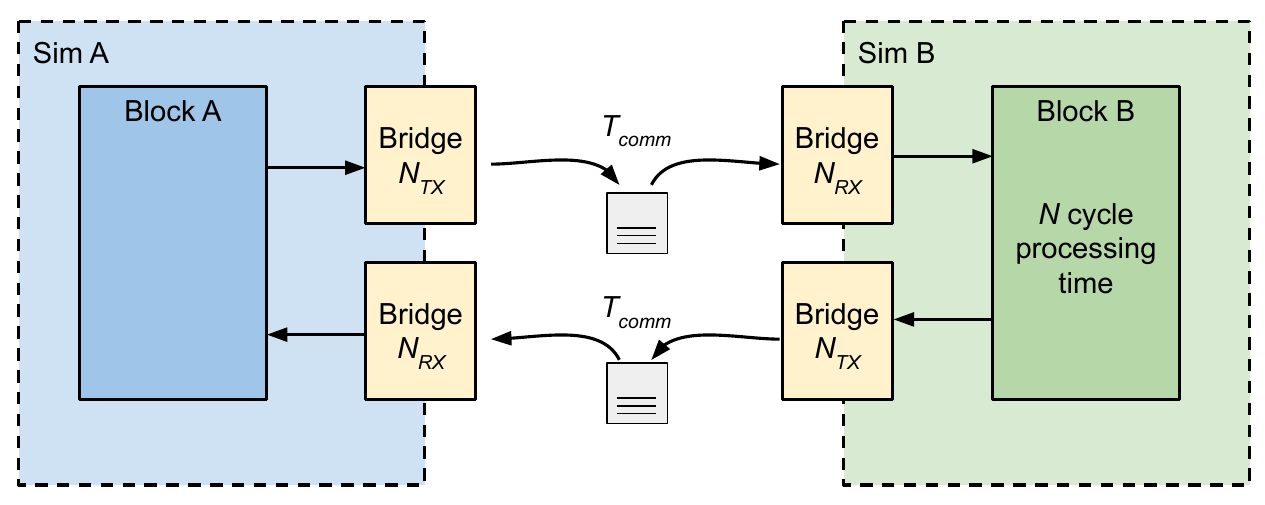}
    \caption{Nonidealities in performance simulation when conveying latency-insensitive interfaces between simulators.}
    \label{fig:performance-sim}
\end{figure}

Figure~\ref{fig:performance-sim} illustrates a minimal version of these issues.  Block A, running in Sim A, sends data to Block B, running in Sim B, which processes the data and sends the result back to Block A.  We assume that each block is driven by a single digital clock, and we make a distinction between the \emph{simulated} and \emph{wall time} rates of those clocks.

The simulated clock rate is the frequency of a clock as measured in simulation; this is the frequency that could be calculated from delay statements in a Verilog model, or could be measured in a waveform viewer.  The wall time clock rate, on the other hand, is the frequency of the clock in real-world time, which is a function of how fast the simulation runs.  For example, if it takes 1~ms to simulate a single clock cycle, the wall time clock rate is 1~kHz.  We denote the simulated clock rates of the two blocks $F_{A,sim}$ and $F_{B,sim}$.  The wall time clock rates are denoted $F_{A,wall}$ and $F_{B,wall}$.

Assuming that Block B takes $N$ clock cycles to process the data it receives, Block A should ideally measure the processing delay (in cycles) to be:
\[
N_{meas,ideal} = NF_{A,sim} / F_{B,sim}
\]
However, unless a simulation is running exactly in real-time, the simulation clock rate will be different from the wall time clock rate.  Hence, the actual processing time measured by Block A is $NF_{A,wall} / F_{B,wall}$.

To correct this, we need to slow down the simulation of either Block A or Block B so that the wall clock rates have the same ratio as the simulated clock rates, i.e. make $F_{A,wall} / F_{B,wall}$ equal to $F_{A,sim} / F_{B,sim}$.  Our Switchboard framework provides a mechanism for doing this by sleeping to fill time as needed to achieve a particular simulation rate.

Counterintuitively, slowing down a block simulation can sometimes speed up a system simulation.  This can happen when there are more simulation processes than hardware threads, if a process that is ``too fast'' is waiting for a slower process.  For example, suppose that Sim A and Sim B must share a hardware thread.  If Sim A is running ``too fast,'' slowing it down will yield more processor time to Sim B, which allows Sim A to see the result of its request faster.  This can lead to an unusual situation where accuracy and speed improve together.

Once the wall clock ratio is corrected, we need to consider the fact that it takes a certain amount of real-world time for data to be conveyed from the bridge module in Block A to the bridge module in Block B, or vice versa.  We denote this time $T_{comm}$.  In addition, the bridge modules themselves may add latency, which we denote $N_{RX}$ for the latency of receiving data from a queue and $N_{TX}$ for the latency of sending data to a queue (both in clock cycles).

Taking all of these nonidealities into account, the processing delay measured by Block A (in clock cycles) is:
\begin{equation*}
\begin{split}
N_{meas,actual} &= NF_{A,wall} / F_{B,wall} \\
                &\quad+ 2T_{comm}F_{A,wall} \\
                &\quad+ \left(N_{RX} + N_{TX}\right)\left(1 + F_{A,wall}/F_{B,wall}\right)
\end{split}
\end{equation*}

This shows that in addition to setting the simulation speeds of Block A and Block B to have the correct ratio, the wall clock rates should be set low enough so that the latency added by inter-block communication is small compared to the ideal measured processing delay, i.e.
\[
F_{A,wall} \ll N_{meas,ideal} / \left(2T_{comm}\right)
\]

The latency added by bridges, $N_{TX}$ and $N_{RX}$, cannot generally be better than one cycle per bridge, because the bridges need to handshake with the block that is sending/receiving data. 
 Hence, this technique for performance measurement should only be used when processing delays are on the order of a few dozen cycles or more.

With these caveats are taken into account, it is possible to achieve reasonably accurate performance estimates, as demonstrated in Section~\ref{sec:millioncore}.

\subsection{Applicability}

We recognize that not all systems are built from modular blocks connected via latency-insensitive interfaces, and hence, not all designs are a good match for Switchboard.  However, constructing large designs this way has important benefits beyond simulation; we use this methodology in our designs for all of those benefits, not just to support simulation.

One of the key benefits of this approach is that hardware blocks, being ``reasonably-sized,'' can run through physical implementation quickly compared to a large flat design.  One could consider chiplet-based design as being an extension of this concept, where modular blocks not only run through physical implementation independently, but are also independently manufactured as physical components, to be integrated in a top-level design on an interposer.

In terms of latency-insensitive channels, we have found that their use eases design integration and reduces the likelihood of bugs, since they eliminate the need to study custom timing diagrams.  They also make it easier to introduce design changes, since one side of a channel can increase or decrease its processing delay without affecting the functionality of the other side.

\section{The Switchboard Framework}
\label{sec:switchboard}

Switchboard is the software framework we developed for putting our modular simulation concept into practice.  It is open source~\cite{switchboard}, released under Apache License 2.0, and is installable from PyPI as \texttt{switchboard-hw}.

Switchboard consists of a fast shared-memory queue implementation (C++), Verilog bridges for connecting latency-insensitive interfaces to queues, Python-based automation for building and running simulations, and a Python API for driving AXI, AXI-Lite, and Universal Memory Interface (UMI) interfaces.

\subsection{Switchboard Packets}

\begin{figure}
    \centering
    \includegraphics[width=\linewidth]{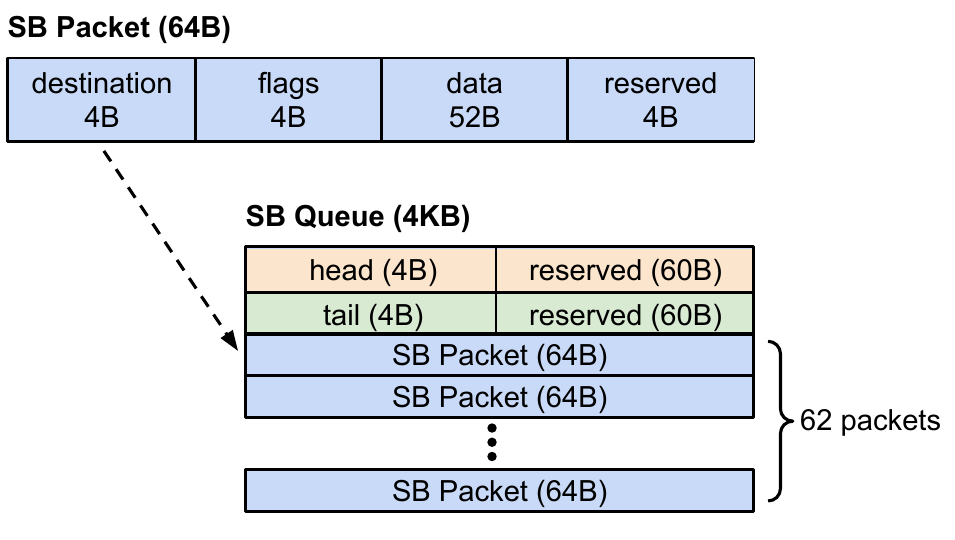}
    \caption{Memory layout for a Switchboard packet and a Switchboard queue.}
    \label{fig:sb-queue-and-packet}
\end{figure}

As described in the previous section, our approach involves conveying latency-insensitive interfaces through SPSC shared-memory queues.  In Switchboard, these queues store 64-byte packets (``SB packets'') consisting of 4 bytes of flags, a 4-byte destination, 52 bytes of data payload, and 4 reserved bytes, as illustrated in Figure~\ref{fig:sb-queue-and-packet}.

The \texttt{flags} and \texttt{destination} fields are intended to allow Switchboard packets to be routed in a generic fashion, without having to inspect the data payload.  At the moment, this feature is mainly used for multiplexing several Switchboard queues on a single TCP port.

\subsection{Queue Implementation}

As shown in Figure~\ref{fig:sb-queue-and-packet}, Switchboard's queue implementation has a standardized memory layout consisting of a 4-byte \texttt{head} where the next packet will be written, a 4-byte \texttt{tail} where the next packet will be read, and 62 slots for SB packets.

To write an SB packet to the queue, the value \texttt{next\_head = (head + 1) \% 62} is computed.  If \texttt{next\_head} equals \texttt{tail}, then the write fails because the queue is full. Otherwise, the SB packet is written to address \texttt{128 + (64 * head)}, and \texttt{head} is set to \texttt{next\_head}.

Reading an SB packet works similarly. If \texttt{tail} equals \texttt{head}, the read fails because the queue is empty. Otherwise, the SB packet is read from address \texttt{128 + (64 * tail)}, and \texttt{tail} is set to \texttt{(tail + 1) \% 62}.

In the full Switchboard queue implementation, cached heads and tails are used to prevent unnecessary cache synchronization between the queue writer and queue reader.  The idea is that if a reader checks \texttt{head} and determines there are multiple packets to read, there is no reason for it to check \texttt{head} again until all of those packets are read (similar idea for a writer and \texttt{tail}).

Other memory optimizations include: placing \texttt{head} and \texttt{tail} in separate cache lines (64B) to prevent false sharing, making an SB packet the size of a cache line, and making a queue the size of a page (4KB).  Acquire/release semantics are used in the implementation to ensure correct queue operation without constraining processor execution more than necessary.

With these optimizations in place, Switchboard queues are fast.  Measurements on one machine\footnote{2.8 GHz quad-core Intel Core i7 with 16 GB 1600 MHz DDR3 memory} indicated a 213~ns round-trip latency and a bandwidth of 27M packets/second, or 1.4 GB/s (per queue).  Since RTL simulations are generally no faster than a few MHz, and are often much slower, Switchboard queues are unlikely to be a bottleneck for modular simulation.

\subsection{Verilog Bridges}

Switchboard provides a number of Verilog modules that serve as bridges between latency-insensitive interfaces and shared memory queues.  These are intended to be attached to ports of a hardware block so that it can communicate with other blocks.

\subsubsection{Low-Level Bridges}

\begin{figure}
    \centering
    \includegraphics[width=1\linewidth]{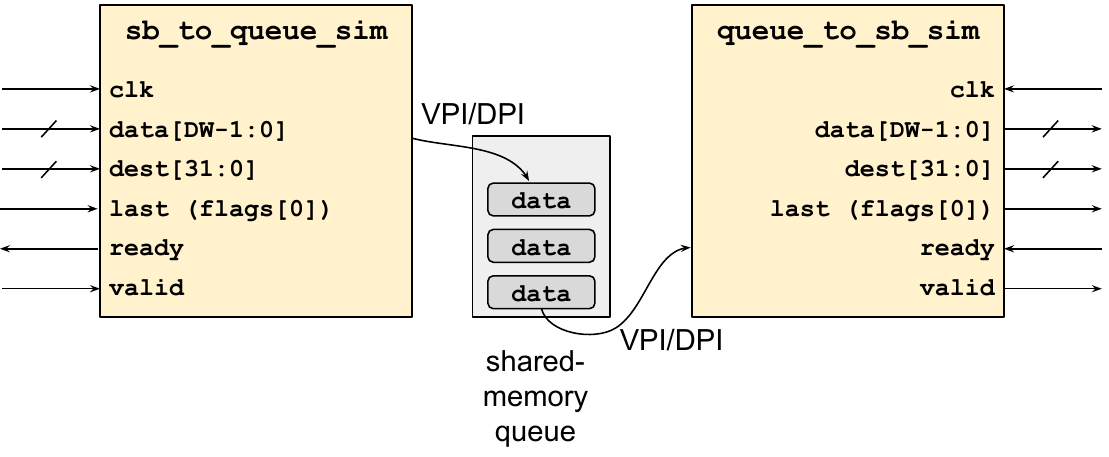}
    \caption{Low-level Verilog bridge interface that conveys SB packets between hardware blocks through a shared-memory queue.}
    \label{fig:verilog-low-level}
\end{figure}

The lowest-level Switchboard bridges operate directly on SB packets.  As illustrated in Figure~\ref{fig:verilog-low-level}, a module called \texttt{sb\_to\_queue\_sim} sends SB packets from a Verilog design to a queue, while a module called \texttt{queue\_to\_sb\_sim} receives SB packets from a queue and drives them into a Verilog design.

Even though Switchboard queues reside in shared memory, they are represented as files in the host file system.  Hence, connecting \texttt{sb\_to\_queue\_sim} and \texttt{queue\_to\_sb\_sim} together is a matter of pointing them to the same file.  This can be done with either a compile-time parameter or at runtime via a module function.

Under the hood, both bridges use Verilog DPI or VPI to interact with shared memory queues.  Which one is used depends on the simulator: Icarus Verilog doesn't support DPI, so VPI is used in that case.  However, DPI is preferred because it is faster.  Switchboard detects whether Icarus Verilog is being used and selects the appropriate interface.

As convenience, we provide macros \texttt{SB\_TO\_QUEUE\_SIM} and \texttt{QUEUE\_TO\_SB\_SIM} for succinctly instantiating low-level bridges.  These macros, and various others, are contained in a Verilog header that ships with Switchboard, \texttt{switchboard.vh}.  The build automation feature described later automatically configures simulators to have this file in their include paths.

\subsubsection{High-Level Bridges}

We provide high-level bridges for conveying several latency-insensitive interfaces: AXI, AXI-Lite, and UMI.  In general, these interfaces consist of multiple latency-insensitive channels, each consisting of a unidirectional (not including handshaking) set of signals.  For example, AXI and AXI-Lite both consist of five latency-insensitive channels: \texttt{AW} (write address), \texttt{W} (write data), \texttt{B} (write response), \texttt{AR} (read address), and \texttt{R} (read response).  UMI consists of two latency-insensitive channels: a request channel and a response channel.  Each latency-insensitive channel is conveyed over a separate Switchboard queue.

\subsubsection{Bridges for FPGA Emulation}

In addition to bridges for RTL simulation, Switchboard also provides bridges for FPGA emulation.  These are intended for a particular type of FPGA system, where user RTL code on the FPGA is given an AXI port by which it can read and write memory on a host system, where shared-memory queues reside.  This is compatible with the development environment for AWS F1 instances, our target for FPGA emulation.

Since all host memory transactions are conveyed through a single AXI port, the Verilog bridges for FPGA emulation are structured differently than the bridges for RTL simulation.  In this case, only one Switchboard bridge is instantiated; it provides one AXI manager interface to interact with host memory, and a parameterized number of input and output ports.  

Switchboard currently ships with two bridges for FPGA emulation: \texttt{sb\_fpga\_queues}, a low-level bridge operating directly on SB packets, and \texttt{umi\_fpga\_queues}, a higher-level bridge for conveying UMI.  As a testament to the portability of our queue specification, the implementations of the queue reader and queue writer are only a few hundred lines of Verilog each.

\subsubsection{Loopback Example}

As a concrete example of how the Verilog bridges can be used, consider a system that receives an SB packet, increments its data field, and transmits the resulting packet.  The Verilog implementation of this system is shown in Listing~\ref{lst:sb-minimal-verilog}.

\begin{lstlisting}[float=t,style={prettyverilog}, caption={Example usage of Switchboard's Verilog bridges.}, label=lst:sb-minimal-verilog, numbers=left]
`include "switchboard.vh"

module testbench (
    `ifdef VERILATOR
        input clk
    `endif
);
    `ifndef VERILATOR
        `SB_CREATE_CLOCK(clk)
    `endif

    localparam integer DW=256;

    `SB_WIRES(to_rtl, DW);
    `QUEUE_TO_SB_SIM(to_rtl, DW, "to_rtl.q");

    `SB_WIRES(from_rtl, DW);
    `SB_TO_QUEUE_SIM(from_rtl, DW, "from_rtl.q");

    assign from_rtl_data = to_rtl_data + 1;
    assign from_rtl_dest = to_rtl_dest;
    assign from_rtl_last = to_rtl_last;
    assign from_rtl_valid = to_rtl_valid;
    assign to_rtl_ready = from_rtl_ready;

    `SB_SETUP_PROBES
endmodule
\end{lstlisting}

The first few lines deal with clock generation in a way that provides broad simulator support.  When Verilator is used, the clock signal is driven from C++ code provided by Switchboard, and hence the \texttt{clk} signal is declared as a top-level input.  For all other simulators, the clock is generated in Verilog with the \texttt{SB\_CREATE\_CLOCK} macro.  In addition to simply toggling the clock, this macro provides hooks for the Switchboard Python interface to adjust the clock frequency at runtime.

Lines 14-18 declare the inbound and outbound SB packet interfaces and associated bridges. 
 \texttt{SB\_WIRES} macros create the interfaces, declaring wires whose names start with the provided prefixes.  The same prefixes are provided to the bridge instantiation macros.  Queue names are specified as the last argument of the bridge macros; if not provided at compile time, they can be configured at runtime through Switchboard's Python interface.

The implementation ends with the \texttt{SB\_SETUP\_PROBES} macro, which sets up waveform probing in a way that can be disabled at either compile time or runtime as an optimization.

\subsection{Python Interface}

Switchboard's Python interface provides build automation, powered by SiliconCompiler~\cite{siliconcompiler}, and an API for driving latency-insensitive interfaces.  As with the Verilog bridges, the Python interface provides both low-level functions for directly sending and receiving SB packets, as well as higher-level functions for issuing read and write transactions over AXI, AXI-Lite, and UMI.

\subsubsection{Low-Level Example}

As a low-level example of Switchboard's Python interface, we demonstrate how it can be used to build a simulator for the previously described increment-and-retransmit module, launch the simulation, send an SB packet into the module, and receive the result.

\begin{python}[caption={Switchboard build automation and API for sending/receiving SB packets.},label={lst:sb-minimal-python},numbers=left,float=t]
import numpy as np
from switchboard import (PySbPacket, PySbTx,
    PySbRx, SbDut)

dut = SbDut()
dut.input('testbench.sv')
dut.build()

tx = PySbTx('to_rtl.q', fresh=True)
rx = PySbRx('from_rtl.q', fresh=True)

dut.simulate()

txp = PySbPacket(destination=0, flags=1,
    data=np.arange(32, dtype=np.uint8))
tx.send(txp)
print(rx.recv())
\end{python}

The Python code that does all of this is shown in Listing~\ref{lst:sb-minimal-python}.  First, a simulator is built with an \texttt{SbDut} object by specifying source files and calling its \texttt{build()} method.  \texttt{SbDut} is a subclass of the SiliconCompiler \texttt{Chip} class, inheriting all of its features for specifying defines, paths, and compilation flags.  By default, Verilator is used as the simulation engine, but this can be changed to Icarus Verilog in the \texttt{SbDut} constructor.

The next step is to create objects for sending and receiving SB packets.  A \texttt{PySbTx} object is created for sending SB packets to a queue called \texttt{to\_rtl.q}; any name could have been used as long as the corresponding Verilog bridge used the same name.  Similarly, a  \texttt{PySbRx} object receives SB packets from a queue called \texttt{from\_rtl.q}, which in turn is driven by the output port of the device under test (DUT).

The simulation is launched on Line 12, and the script interacts with the running simulation in the lines that follow.  Python code creates an SB packet and transmits it to the DUT with the \texttt{PySbTx.send()} method; the SB packet returned by the DUT is received with the \texttt{PySbRx.recv()} method.  The \texttt{data} field in SB packets is represented with a NumPy~\cite{numpy} array, which provides a convenient and efficient mechanism for passing arrays around in Python.

\subsubsection{High-Level Example}

In addition to providing functions for directly sending and receiving SB packets, Switchboard provides functions for driving three higher-level interfaces: AXI, AXI-Lite, and UMI.  In all three cases, low-level \texttt{send()} and \texttt{recv()} methods are replaced with \texttt{read()} and \texttt{write()} methods that operate in an address space.

\begin{python}[caption={Example of Switchboard's API for driving higher-level latency-insensitive interfaces.},label={lst:sb-axi-python},numbers=left,float=t]
import numpy as np
from switchboard import SbDut, AxiTxRx

dut = SbDut()

# define RTL sources...

axi = AxiTxRx('axi', data_width=...,
    addr_width=..., id_width=...)

dut.simulate()

axi.write(0x1234, np.arange(42, dtype=np.uint16))
x = axi.read(0x1234, 42, dtype=np.uint16)
\end{python}

As an example, consider the implementation of a system where a Python script issues reads and writes to an AXI memory implemented in Verilog.  The corresponding Python code is shown in Listing~\ref{lst:sb-axi-python}; for brevity, the definition of source files is omitted.

Lines 8-9 instantiate an \texttt{AxiTxRx} object that will be used to issue AXI reads and writes.  Internally, this sets up 5 queues, one for each of the latency-insensitive AXI channels.  The first argument of the constructor specifies the prefix used for naming these queues; in this example, the queues will be named \texttt{axi-aw.q}, \texttt{axi-w.q}, \texttt{axi-ar.q}, etc.

After the simulation is launched, the \texttt{AxiTxRx} object is used to interact with the AXI memory.  The \texttt{AxiTxRx.write()} method writes a NumPy scalar or array to a given starting address, and similarly the \texttt{AxiTxRx.read()} method reads a NumPy scalar or array from a given starting address.  Under the hood, Switchboard generates bursts and/or multiple transactions if needed, performs alignment, and generates write strobes.  The result is that the users can interact with an address space without having to know about the width of AXI data buses or consider alignment.

A similar abstraction is provided for UMI through the \texttt{UmiTxRx} class, with the addition of an \texttt{atomic()} method that provides a mechanism for generating UMI atomic transactions.

\subsection{Autowrap Feature}

As the complexity of a hardware block grows, it can become cumbersome to keep track of queue names and maintain a Verilog wrapper.  For this reason, we provide a feature that automatically generates Verilog wrapper code for a DUT and sets up transceivers to interact with its interfaces.  The feature is enabled by setting \texttt{autowrap=True} when instantiating \texttt{SbDut}.

\begin{python}[caption={Example of Switchboard's \texttt{autowrap} feature.},label={lst:sb-autowrap},numbers=left,float=t]
import numpy as np
from switchboard import SbDut, AxiTxRx

dw, aw, idw = 32, 13, 8

interfaces = {'s_axi': dict(type='axi', dw=dw,
    aw=aw, idw=idw, direction='subordinate')}

resets = [dict(name='rst', delay=8)]

parameters = dict(DATA_WIDTH=dw,
    ADDR_WIDTH=aw, ID_WIDTH=idw)

dut = SbDut('axi_ram', autowrap=True,
    interfaces=interfaces, resets=resets,
    parameters=parameters)

# define RTL sources...

dut.simulate()

axi = dut.intfs['s_axi']

axi.write(0x1234, np.arange(42, dtype=np.uint16))
x = axi.read(0x1234, 42, dtype=np.uint16)
\end{python}

As an example, Listing~\ref{lst:sb-autowrap} shows a re-write of the previous AXI example using autowrap.  There are two key differences with respect to the previous version: first, the \texttt{SbDut} constructor is given additional arguments that specify how the Verilog module should be wrapped.  Second, rather than explicitly creating transceiver objects, they are automatically created when simulation starts, and made available through the \texttt{SbDut} object.

This functionality is primarily controlled via the \texttt{interfaces} argument of the \texttt{SbDut} constructor. This argument is a dictionary mapping the signal prefix of an interface to a dictionary of properties for that interface.  In this example, there is one interface whose signals start with the prefix \texttt{s\_axi}; we specify the directionality and signal widths of this interface as properties.  Although not shown here, some types of signals that are not part of latency-insensitive interfaces can also be autowrapped, such as signals that need to be tied off to constants.

When the autowrap feature is used, the simulation launch process sets up transceivers for the latency-insensitive interfaces specified via the \texttt{interfaces} argument.  These transceivers are made available through the \texttt{SbDut.intfs} property, which is a dictionary keyed by signal prefix (i.e., the same keys as the \texttt{interfaces} dictionary).  In this example, the \texttt{intfs} dictionary has a single entry corresponding to the \texttt{s\_axi} interface.  Other interface types supported by autowrap include AXI-Lite, UMI, and low-level SB packets.

\subsection{Network Construction}

Up until this point, our discussion has been limited to a single hardware block.  This can be useful in its own right for Python-based block-level testing.  However, the purpose of Switchboard is ultimately to support the construction of large simulations from multiple hardware blocks.  This is accomplished with Switchboard's \texttt{SbNetwork} abstraction.

\subsubsection{SbNetwork}

The general idea of \texttt{SbNetwork} is to create multiple \texttt{SbDut} objects, one for each unique hardware block, and instantiate these objects one or more times in a network, with connections established between pairs of latency-insensitive ports on block instances.  Some ports are made external, becoming ports on the network itself.

\begin{figure}
    \centering
    \includegraphics[width=1\linewidth]{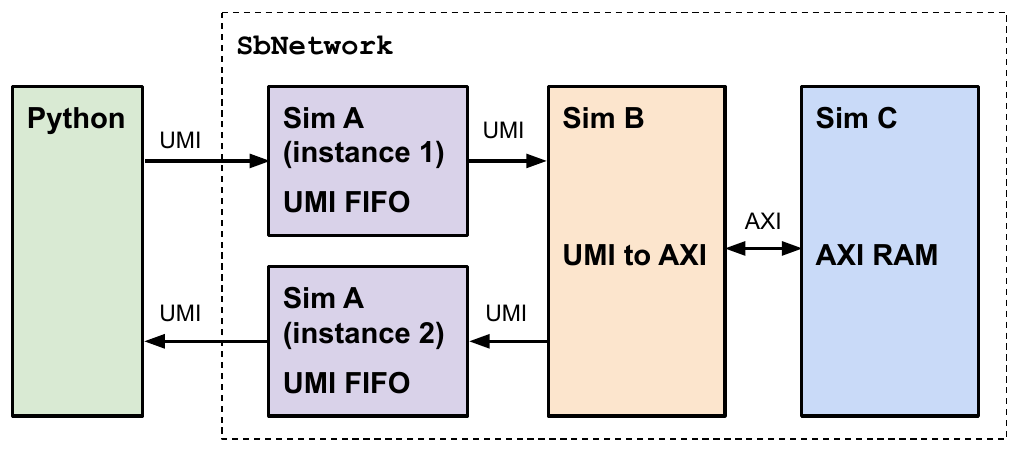}
    \caption{Example of the \texttt{SbNetwork} abstraction, showing how a system can be modeled by instantiating a prebuilt simulator for each hardware block and wiring them together.}
    \label{fig:sb-network-example}
\end{figure}

A small example is shown in Figure~\ref{fig:sb-network-example}, where a Python script interacts with an AXI memory that sits behind a UMI-to-AXI converter and UMI FIFOs.  There are three unique blocks: the UMI FIFO, the UMI-to-AXI converter, and the AXI memory.  The FIFO is instantiated twice in the network, and the other two modules are instantiated once.

\begin{python}[caption={Example of the \texttt{SbNetwork} abstraction.},label={lst:sb-network},numbers=left,float=t]
from switchboard import SbNetwork

net = SbNetwork()

umi_fifo, umi2axil, axil_ram = ... # type: SbDut

umi_fifo_i = net.instantiate(umi_fifo)
umi_fifo_o = net.instantiate(umi_fifo)
umi2axil_i = net.instantiate(umi2axil)
axil_ram_i = net.instantiate(axil_ram)

net.connect(umi_fifo_i.umi_out,
    umi2axil_i.udev_req)
net.connect(umi_fifo_o.umi_in,
    umi2axil_i.udev_resp)
net.connect(umi2axil_i.axi, axil_ram_i.s_axil)

net.external(umi_fifo_i.umi_in, txrx='umi')
net.external(umi_fifo_o.umi_out, txrx='umi')

net.build()
net.simulate()

umi = net.intfs['umi']
# umi.write(), umi.read(), etc.
\end{python}

The corresponding Python implementation for this network is in Listing~\ref{lst:sb-network}.  After \texttt{SbDut} objects are created, they are instantiated one or more times in the network via the \texttt{SbNetwork.instantiate()} method.  This method returns an \texttt{SbInst} object representing a specific instance, with properties named after the instance's interfaces.  For example, the UMI FIFO block has interfaces named \texttt{umi\_in} and \texttt{umi\_out}, so the instance \texttt{umi\_fifo\_i} has properties with those names.  This feature is put to use in lines 12-16, which set up connections between hardware blocks in the network.  The \texttt{SbNetwork.connect()} method is used to create each connection; it accepts two arguments, corresponding to the two interfaces to be connected.

On Lines 18-19, two interfaces are marked as external to the network, to be made available to the Python script for interaction.  Building the network causes simulators to be built for each unique hardware block (three in this case), while simulating the network causes a simulator to be launched for each block instance (four in this case).  There is no reference to specific queue names in this Python code; queues are automatically assigned, and queue names are passed to simulations at runtime.

Once the network is running, interfaces marked external are available through the \texttt{SbNetwork.intfs} property, as with the \texttt{SbDut.intfs} property when the autowrap feature is used.  When the Python script exits, all running simulations are automatically shut down, and the queues allocated are deleted.

\subsubsection{Single-Netlist Simulation}

It is sometimes desirable to temporarily disable Switchboard-based communication and model an entire system in a single RTL simulation.  For example, in the manycore experiment (Section~\ref{sec:millioncore}) we needed to gather ground truth performance measurements, unaffected by queue latency, simulators running at different speeds, etc.  Switchboard supports this type of use case via the \texttt{single\_netlist=True} option in the \texttt{SbNetwork} constructor.

When single-netlist simulation is enabled, \texttt{SbNetwork.build()} builds a single simulator for the entire network; the simulator is launched with \texttt{SbNetwork.simulate()}. 
 Although this behavior selected with a single flag, its implementation is quite different from the distributed simulation implementation, as Switchboard generates Verilog code that implements the network.

\subsubsection{Network-of-Networks}

When hardware blocks are relatively small, it may not be efficient to run a separate simulator process for each hardware block, due to the overhead of context switches.  One would expect that the best simulation performance is achieved when there is one simulation process per CPU core, making use of all CPU resources, but avoiding the need for context switches.

In support of this type of optimization, we provide a feature that enables subsets of a large system to each run as a single-netlist simulation.  We refer to this as simulating a ``network-of-networks.''

Building a network of networks is straightforward: create an \texttt{SbNetwork} object with \texttt{single\_netlist=True} and \texttt{instantiate()} in another \texttt{SbNetwork}.  Just as with an \texttt{SbDut} object, an \texttt{SbNetwork} object may be instantiated multiple times.

\subsubsection{Bridging over Ethernet}

\begin{python}[caption={Example of bridging Switchboard connections over Ethernet.},label={lst:sb-tcp},numbers=left,float=t]
from switchboard import SbNetwork, TcpIntf

net = SbNetwork()
dut = ... # SbDut()

dutA = net.instantiate(dut)
dutB = net.instantiate(dut)

net.connect(dutA.umi_in, TcpIntf(
    port=5555, mode='server'))

net.connect(dutB.umi_out, TcpIntf(
    port=5555, host='1.2.3.4', mode='client'))
\end{python}

For simulating very large networks, it may be necessary to use multiple host machines that are bridged over Ethernet.  This is straightforward to set up in an \texttt{SbNetwork} by connecting an interface on a hardware block to a \texttt{TcpIntf} object, as illustrated in Listing~\ref{lst:sb-tcp}.

As shown in the example code, the port used for a bridge is user-specified, and each bridge can be run as either a TCP server or TCP client, depending on whichever is more convenient (although the opposite side of the bridge must use the opposite mode).  When running a simulation across cloud instances and driving the test stimulus from a local machine, it is convenient to avoid using server mode on the local machine, since that often requires local router settings to be altered.

Under the hood, Switchboard launches TCP-based bridges when \texttt{SbNetwork.simulate()} is called, and care is taken to make sure that these processes are shut down when the network shuts down.  There is no restriction as to the ordering of launching clients and servers.

\subsection{Mixed-Signal Simulation}

We have started to explore support for SPICE models in Switchboard.  While analog circuit interfaces are not latency-insensitive, the idea of being able to incorporate SPICE models is consistent with the framework's goal of being able to wire together many different types of models.  

\begin{figure}
    \centering
  \includegraphics[width=0.7\linewidth]{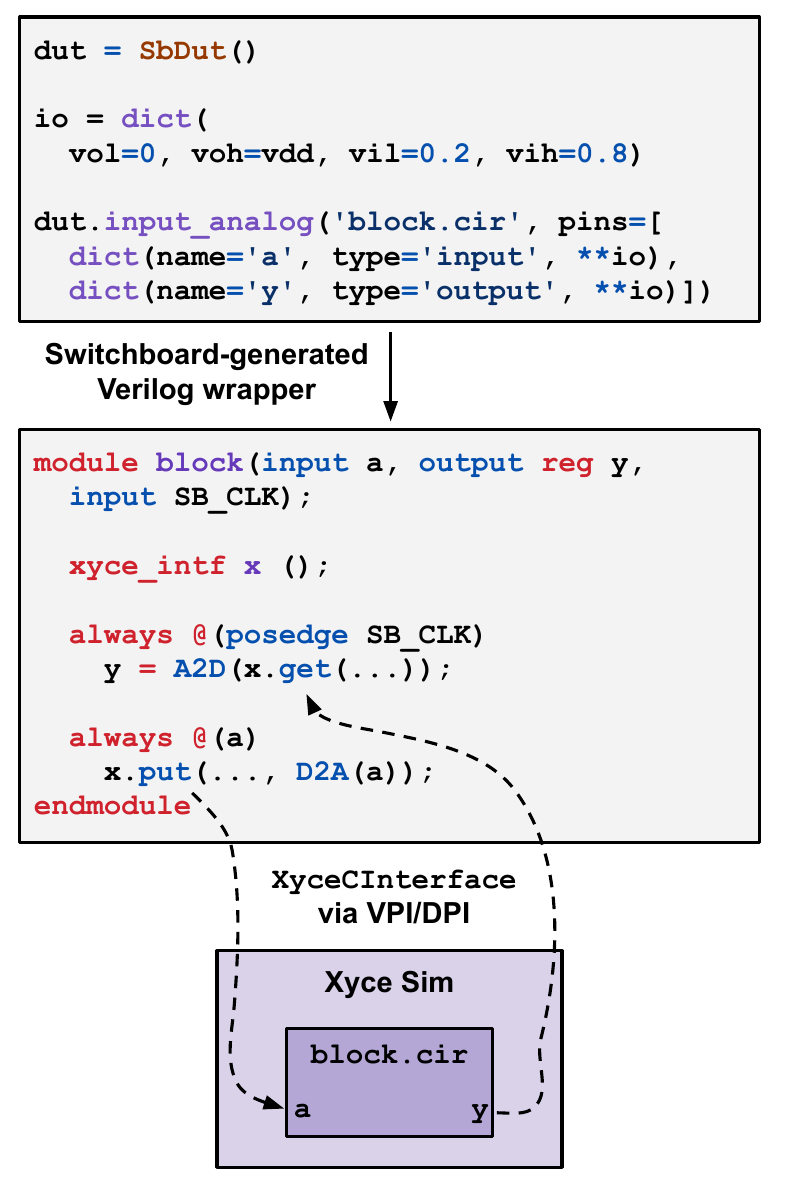}
    \caption{Switchboard provides a mechanism to instantiate SPICE models in Verilog via a generated wrapper that interacts with a running Xyce simulation.}
    \label{fig:mixed-signal}
\end{figure}

Our current implementation supports direct instantiation of SPICE subcircuits in user RTL.  As illustrated in Figure~\ref{fig:mixed-signal}, SPICE model sources are specified as \texttt{SbDut} input files in a manner similar to RTL sources, with an additional argument that indicates how digital signals should be converted to analog inputs (D2A) and how analog outputs should be converted to digital signals (A2D). 
 For each SPICE model, Switchboard generates a Verilog wrapper that performs the D2A and A2D conversions and uses VPI or DPI to write values to and read values from a running SPICE simulation.

SPICE simulations are run with Xyce~\cite{keiter2023xyce}, using \texttt{XyceCInterface} to make the simulation controllable from RTL simulation.  Analog inputs are represented as piecewise-linear waveforms with breakpoints defined by value changes in the digital simulation.  Analog outputs, on the other hand, are oversampled, with time in both the RTL and SPICE simulations advancing in increments of the oversampling period.  In the future, we hope to use Xyce's ability to simulate until an analog threshold is crossed as a more efficient alternative to oversampling.

\section{Applications}
\label{sec:applications}

In this section, we discuss two rather different applications of Switchboard.  The first was to power a web app where users could interact with different combinations of chiplets on a virtual substrate, while the second was to simulate an array of one million RISC-V cores in a wafer-scale system.  The underlying theme is that Switchboard is fast, scalable, and flexible.

In the experiments described below, the local machine used for comparison with cloud simulation had a 2.8 GHz quad-core Intel Core i7 with 16 GB 1600 MHz DDR3 memory.

\subsection{Drag-and-Drop Chiplets}

\begin{figure}
    \centering
    \includegraphics[width=0.5\linewidth]{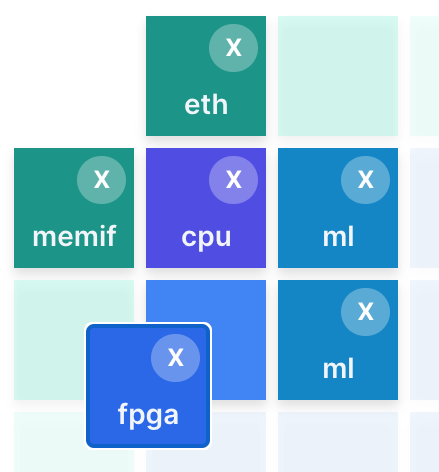}
    \caption{User interface of the Switchboard-powered web app where users could drag and drop chiplets from a catalog onto a virtual silicon substrate.}
    \label{fig:web-app}
\end{figure}

Switchboard was originally created as the underlying technology for the Zero ASIC emulation web app~\cite{webapp}.  As illustrated in Figure~\ref{fig:web-app}, web app users could drag and drop chiplets from a catalog onto a virtual silicon substrate to create a custom system.  The chiplet catalog consisted of five parts: a RISC-V CPU, an ML accelerator, an FPGA, a DRAM interface, and an Ethernet MAC.  In line with our design philosophy, the CPU, FPGA, and ML chiplets were ``reasonably sized'' at 2x2 mm, targeting a 12 nm process.

Through the web app, users could boot Linux on the CPU chiplet from an image loaded through the DRAM interface, and SSH into the Linux system through the Ethernet chiplet.  From that point, custom applications could be run making use of any of the chiplets in the catalog.

The CPU, ML, and FPGA chiplets were represented using their respective RTL implementations to ensure that the models were accurate and up-to-date.  Since RTL simulations of the CPU and ML chiplets would have been too slow to support booting Linux or running ML models, we sought to implement these chiplets on FPGAs.

In the context of a web app, any FPGAs used needed to be accessible over the internet by a pool of users.  We chose AWS F1 instances for this purpose to avoid having to build, grow, and maintain our own FPGA infrastructure.  F1 instances can be configured with 1-8 FPGAs, each containing a Xilinx VU9P FPGA.  The FPGAs are connected to a host system over PCIe.

\begin{figure}
    \centering
    \includegraphics[width=\linewidth]{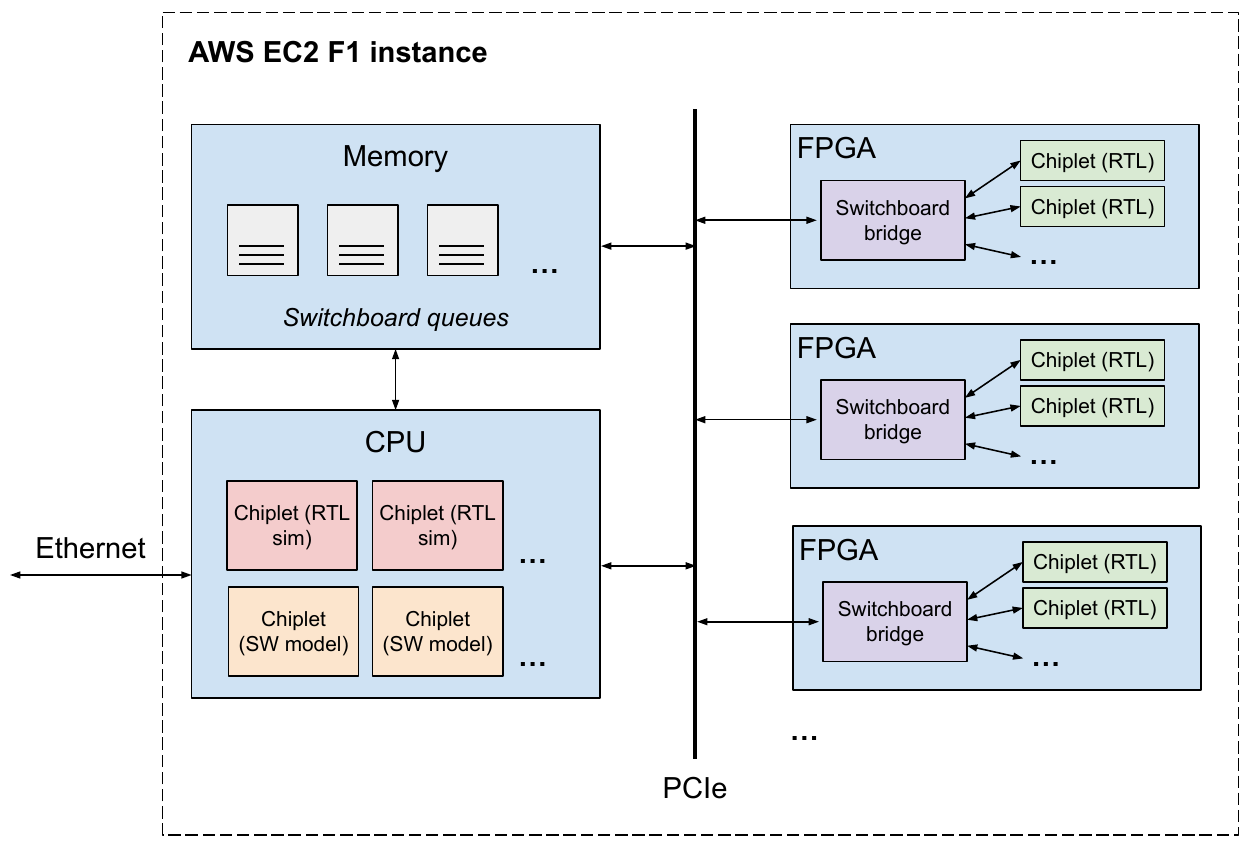}
    \caption{Simulation architecture for the web app, where chiplets can be represented by RTL simulation, RTL implemented on FPGAs, or SW models.}
    \label{fig:web-app-arch}
\end{figure}

The overall simulation architecture for the web app is illustrated in Figure~\ref{fig:web-app-arch}.  Each FPGA contains one Switchboard bridge, connected to one or more chiplet designs, and these bridges interact with shared-memory queues in host memory.  All FPGA bitstreams were prebuilt and loaded onto FPGAs as needed to implement the system defined by a user.  Chiplets not implemented on FPGAs were simulated on the host processor, either with SW models (C++) or via RTL simulation using Verilog simulation bridges.  These simulations were also prebuilt.

The result was that simulations of custom systems could be started quickly because there was no need to build designs at runtime; starting a simulation was just a matter of launching prebuilt simulators representing the constituent hardware blocks.

Our web app consisted of two FPGA images, one containing an instance of the CPU chiplet and an Ethernet chiplet, and the other consisting of the ML chiplet.  As shown in Table~\ref{tab:fpga-vs-sim}, these images produced speedups of almost four orders of magnitude compared to simulations of the same RTL run on a local machine.  While it is well-known that FPGA emulation is much faster than RTL simulation, what was unique here was the ability to take advantage of FPGA-level performance in a heterogeneous simulation.

\begin{table}[t]
    \centering
    \renewcommand{\arraystretch}{1.3}
    \caption{Comparison of FPGA emulation and RTL simulation performance}
    \label{tab:fpga-vs-sim}
    \begin{tabular}{|c|c|c|c|}
        \hline
         Image & FPGA Clock & RTL Sim Rate & Speedup \\
         \hline\hline
         CPU+ETH & 125 MHz & 14 kHz & 8,900x \\
         \hline
         ML & 31.25 MHz & 4.3 kHz & 7,300x \\
         \hline
    \end{tabular}
\end{table}

This speedup was critical because we wanted to run interactive, high-level tasks in the web app.  As demonstrated by the results in Table~\ref{tab:interactive-tasks}, we achieved this goal: booting Linux took a few minutes, and a convolutional neural network (CNN) image classifier could run in under half a minute.  Although the CNN inference speed is certainly slower than the target for physical silicon, it was acceptably fast for an interactive application, especially given that feedback was printed as the CNN ran.

\begin{table}[!t]
    \centering
    \renewcommand{\arraystretch}{1.3}
    \caption{Measured duration of high-level tasks}
    \label{tab:interactive-tasks}
    \begin{tabular}{|c|c|c|}
        \hline
         Task & Time in Web App & Projected Sim Time \\
         \hline\hline
         Boot Linux & 1m 51s & 11d \\
         \hline
         ML Inference & 19s & 1.6d \\
         \hline
    \end{tabular}
\end{table}

Building FPGA images took 5h 40m for the CPU+ETH combination and 7h 43m for the ML chiplet; this includes the time needed to postprocess FPGA design checkpoints into Amazon FPGA Images (AFIs).  As shown in Table~\ref{tab:resource-util}, both implementations fit comfortably on F1 FPGAs.  The ML chiplet design was more taxing because its arithmetic blocks did not map well to FPGA DSPs.

\begin{table}[t]
    \centering
    \renewcommand{\arraystretch}{1.3}
    \caption{Resource utilization of FPGA images}
    \label{tab:resource-util}
    \begin{tabular}{|c|c|c|c|c|c|}
        \hline
         Image & LUT & FF & BRAM & URAM & DSP \\
         \hline\hline
         CPU+ETH & 19\% & 7\% & 11\% & 0.4\% & 2\% \\
         \hline
         ML & 59\% & 9\% & 0\% & 0\% & 3\% \\
         \hline
    \end{tabular}
\end{table}

Unlike the CPU, ML, and Ethernet chiplets, the FPGA chiplet was represented with RTL simulation.  This was due to a constraint of cloud FPGA development: to prevent physical damage, AFIs are prohibited from containing combinational loops.  As with any FPGA, it was possible to implement a combinational loop on the FPGA chiplet, and hence we could not pass this check when attempting to build its AFI.  Nonetheless, we synthesized the design and found that it could have been represented with about 100-200 F1 LUTs per LUT in the FPGA chiplet.  This suggests that FPGA architectures of up to a few thousand LUTs could be implemented on an F1-sized FPGA (for example, if running an on-premise FPGA cluster without the combinational loop ban).

Despite the slow speed of RTL simulation of the FPGA chiplet (159 Hz), it was possible to interact with the chiplet by reading and writing its GPIO buses, which took 2-4s per interaction.  This enabled web app users to implement simple RTL designs on the FPGA chiplet and test them out through manual interaction.

The DRAM interface was modeled directly in C++ using Switchboard, since there was no corresponding RTL implementation.  This resulted in a fast model with a write bandwidth of 143 MB/s and a read bandwidth of 130 MB/s, including protocol overhead and the overhead of initiating reads and writes from Python.

The overall takeaway from this application is that Switchboard is flexible, allowing hardware blocks to be represented with RTL simulation, RTL implemented on FPGAs, or SW models, depending on what is most appropriate for each block.  In addition, Switchboard supports a high level of emulation performance that enables interactive high-level tasks to be run on complex hardware systems.

\subsection{Million-Core Simulation}
\label{sec:millioncore}

The previous application demonstrated modest scaling, with the capacity to represent a few dozen hardware blocks.  We now examine an application with extreme scaling: simulating an array of one million RISC-V cores in the cloud, as a proof-of-concept wafer-scale simulation.

\subsubsection{Architecture}

The million-core simulation was run using AWS Elastic Container Service (ECS), which supports the creation of clusters that run container-based tasks.  We chose ECS over direct use of EC2 because it avoided the need to manage individual cloud instances, which can be cumbersome when working with many instances.  In addition, deploying environments with containers is faster and more flexible than working with Amazon Machine Images (AMIs), since containers can be built locally and updated incrementally.

ECS tasks can be dispatched either to EC2 instances or, using Fargate, to anonymized cloud compute resources specified by a virtual CPU (vCPU) count and an amount of memory.   Fargate was chosen as the ECS capacity provider because it avoids the need to keep cloud resources running when not in use, helping to reduce costs when working with a large number of tasks.

\begin{figure}
    \centering
    \includegraphics[width=1\linewidth]{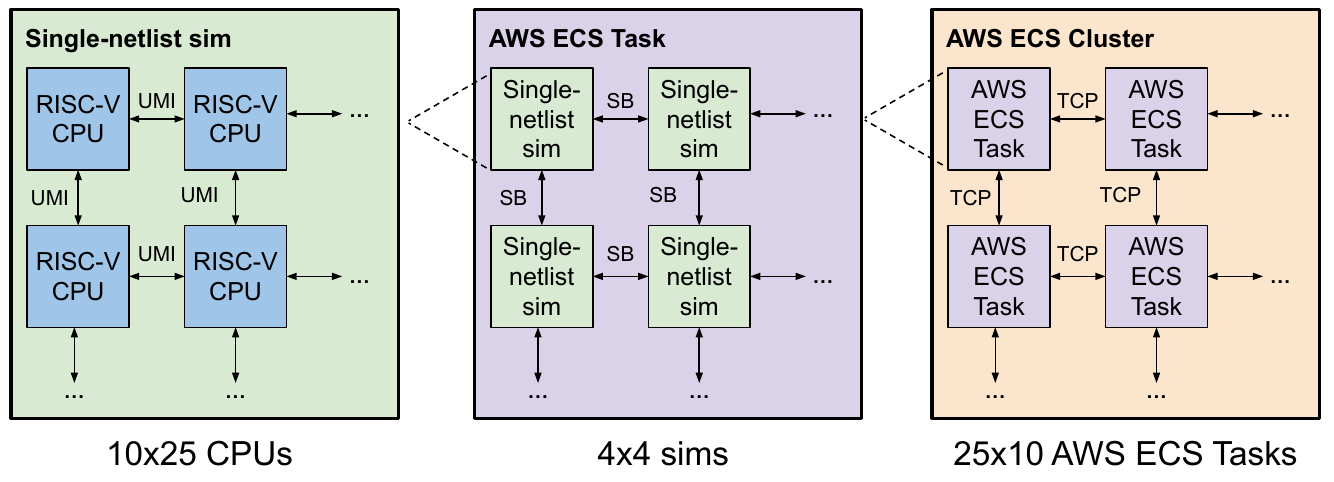}
    \caption{Architecture of the million-core simulation.}
    \label{fig:large-sim-arch}
\end{figure}

The overall simulation architecture is illustrated in Figure~\ref{fig:large-sim-arch}.  A single-netlist simulation of a 10x25 array of RISC-V processors was built, and a 4x4 array of those simulations was run in a Fargate-based AWS ECS task, connected through Switchboard queues.  These tasks were in turn formed into a 25x10 array in an ECS cluster and connected with Switchboard TCP bridges, representing a total of one million RISC-V cores.

For each ECS task, we used 16 vCPUs, the maximum allowed with Fargate, and 32 GB RAM.  In general, for high performance it is desirable to make communication as tightly-coupled as possible, and hence we would rather use a smaller number of ECS tasks with the maximum number of vCPUs, as opposed to a larger number of ECS tasks with fewer vCPUs.  Considering that we used a 25x10 array of ECS tasks, the total number of vCPUs involved in the simulation was 4,000.

The RISC-V processor used in the array was PicoRV32~\cite{picorv32}, a small, open-source 32-bit implementation.  Each processor was wrapped with additional hardware that enabled it to communicate with its neighbors to the north, east, south, and west.  For a sense of scale, this unit cell could be implemented with about 20k gates (198k transistors) in the ASAP7~\cite{asap7} open-source PDK.

\begin{figure}
    \centering
    \includegraphics[width=0.9\linewidth]{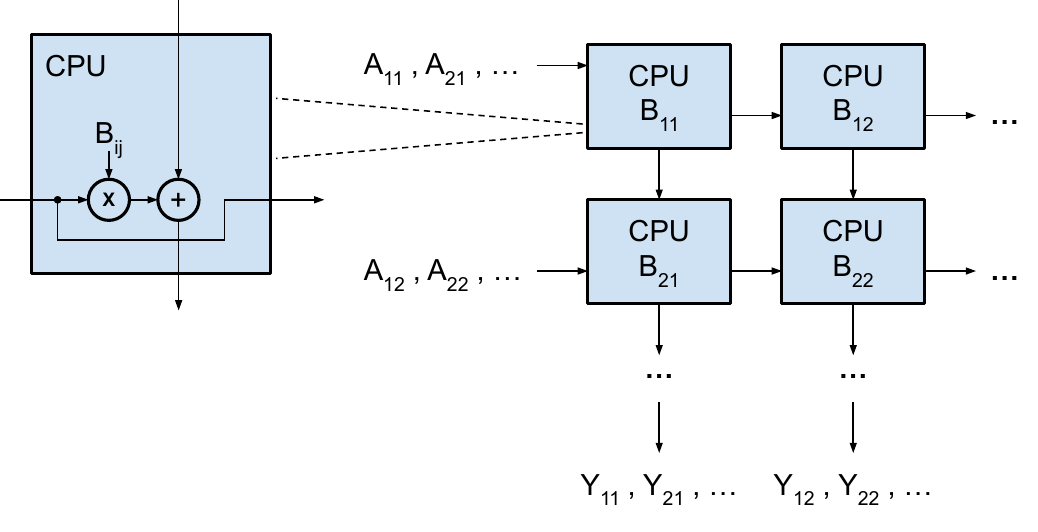}
    \caption{The RISC-V processor array was used to compute matrix multiplications in a distributed fashion.}
    \label{fig:matmul}
\end{figure}

We used the RISC-V processor array to implement the matrix multiplication $Y = A \times B$ as illustrated in Figure~\ref{fig:matmul}.  Each CPU stored an element of $B$, performing one multiplication and one addition for each packet it received.  The transposed rows of $A$ flowed into the array on its west edge and moved eastward, while partial sums flowed from north to south, with the rows of $Y$ appearing on the south edge.

\subsubsection{Building and Running the Simulation}

Only one RTL simulation needed to be built to run the million-core simulation: the single-netlist simulation of a 10\texttimes 25 grid of PicoRV32 processors.  This modestly-sized simulation took 3m 26s to build on a local machine, and the resulting simulation binary was uploaded to AWS Simple Storage Service (S3) where it could be accessed by ECS tasks.

Each ECS task ran 16 copies of this simulation in a 4\texttimes 4 grid, and the ECS tasks themselves were arranged in a 25\texttimes 10 grid, resulting in a total of 4,000 copies of the simulation.  The simulation took 10m 54s to run a single 1000\texttimes 1000 by 1000\texttimes 1000 matrix multiplication, with the timing breakdown shown in Table~\ref{tab:timing-breakdown}.  Including build time, it was possible to go from RTL to a behavioral simulation result in under 15m.

\begin{table}
    \centering
    \renewcommand{\arraystretch}{1.3}
    \caption{Timing breakdown for the million-core simulation}
    \label{tab:timing-breakdown}
    \begin{tabular}{|c|c|c|}
        \hline
        Name & Time & Percentage \\
        \hline\hline
        Launch 250 ECS tasks & 2m 30s & 23\% \\
        \hline
        Wait for ECS tasks to boot & 1m 20s & 12\% \\
        \hline
        Run simulation & 7m 4s & 65\% \\
        \hline
        Total & 10m 54s & 100\% \\
        \hline
    \end{tabular}
\end{table}

In the timing breakdown, we observe that a fair amount of time was taken to launch ECS tasks.  This appears to be due to AWS rate limiting of the speed at which ECS tasks could be created.  The time taken for ECS tasks to boot was also noticeable; this was independent of the number of tasks and instead related to the size of the Docker image used in the task definition (1.5 GB).  As a future optimization, ECS tasks could be run in a warm pool of EC2 instances with Docker image caching to reduce this overhead, although this would increase cost.

The million-core simulation was run in the Northern California AWS region (\texttt{us-west-1}), where the cost of Fargate resources was \$0.04656/vCPU/hr and \$0.00511/GB memory/hr.  Hence, the cost of running the million-core simulation was \$41.26.  Although not trivial, this is likely reasonable within the context of a budget for a wafer-scale project.  In normalized terms, the cost of simulation was approximately 0.16 cents per megagate-megacycle, a value that could be used in back-of-the-envelope cost estimations for other systems.

\subsubsection{Comparison to Multithreaded Verilator}

To explore the value of using Switchboard, we consider an alternative: building and running a single multithreaded (MT) Verilator simulator for the entire RISC-V array on a large machine.  For this study, the large machine used was an AWS EC2 instance of \texttt{c6a.48xlarge}, which provides 192 vCPUs and 384 GB RAM.

\begin{figure}
    \centering
    \includegraphics[width=\linewidth]{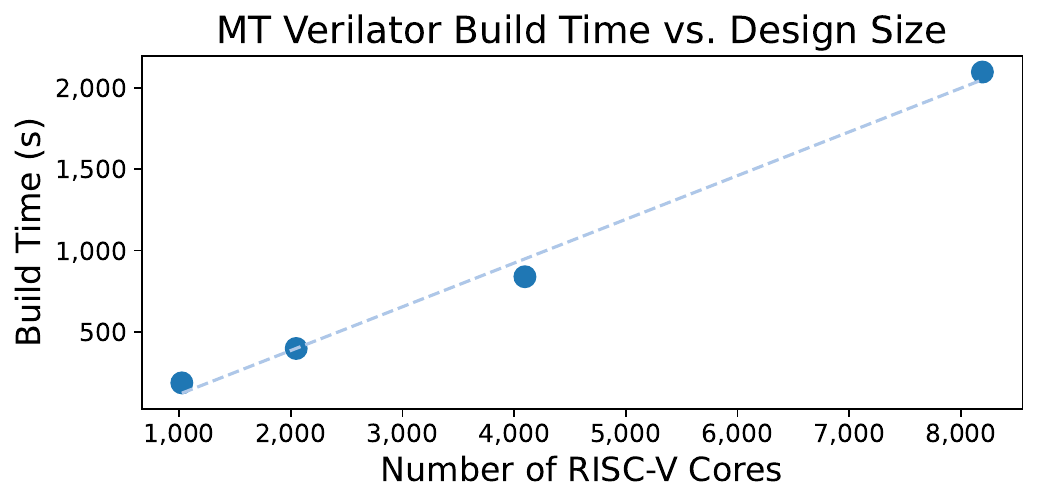}
    \caption{Build time for different design sizes when using multithreaded Verilator instead of Switchboard.}
    \label{fig:build-time}
\end{figure}

To start, we examined build time, scaling up the size of the RISC-V array in factors of two while recording the time needed to build an MT Verilator simulator.  The result is shown in Figure~\ref{fig:build-time}, demonstrating a fairly linear trend.  We stopped at 8,192 cores, where the build time had grown to around half an hour.

\begin{figure}
    \centering
    \includegraphics[width=\linewidth]{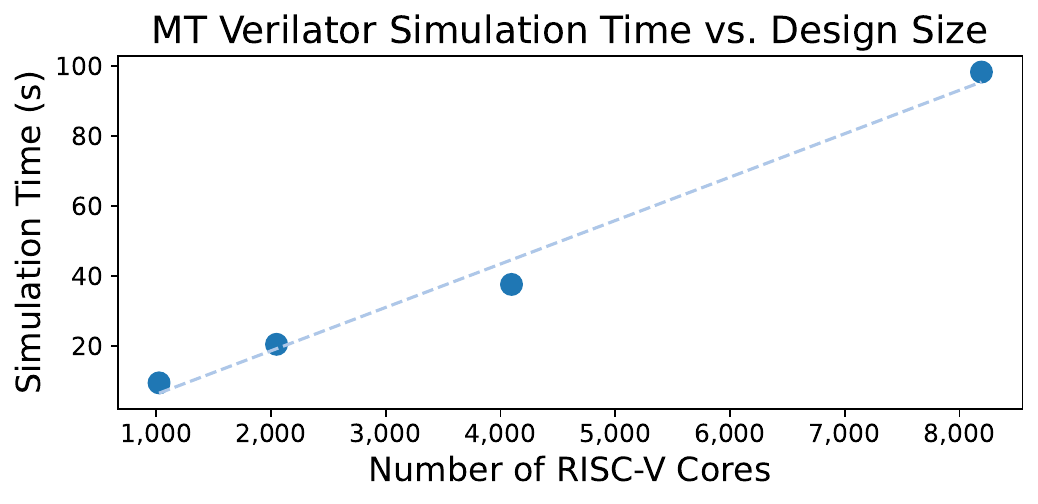}
    \caption{Time to simulate one matrix multiplication for different design sizes when using multithreaded Verilator instead of Switchboard.}
    \label{fig:mt-verilator-sim-time}
\end{figure}

Although it is unknown whether a 1M core simulation could be built in any amount of time on this machine, we can arrive at a very rough estimate by extrapolating the trend line.  This is clearly problematic because we are extrapolating over two orders of magnitude, however it is still useful to get a sense of how long such a build would take.  The extrapolation yields an estimated million-core build time of 3.1 days, with is 1,300x slower than with Switchboard.  Given that other work \cite{metro-mpi} reported a 2.9 day build time for a smaller design (10B transistors), this is likely an underestimate.

For each design size in the experiment, we measured how long it took to simulate a single matrix multiplication, as shown in Figure~\ref{fig:mt-verilator-sim-time}.  The trend is linear, though we suspect it would become at least quadratic at larger design sizes.  The reason is that the number of threads for each design size was selected for best performance, with the 1,024-core design using 32 threads and the 8,192-core design using 128.  Since the machine itself only has 192 vCPUs, the thread count could not be productively increased much more for larger designs.  Once all vCPUs are active running simulation threads, the simulation rate would decrease at least linearly with increasing design size, while the number of cycles to simulate would increase linearly, leading to a quadratic trend.  

As a result, we believe that extrapolating the linear trend line of simulation time should yield a conservative estimate, in the sense that it would be biased against Switchboard.  Performing that extrapolation, we predict that the MT Verilator simulation time for the million-core array would be at least 3h 27m, which is 19x slower than with Switchboard.  This seems plausible, given that the Switchboard-based simulation was run on a cluster whose vCPU count was greater than that of the single-machine simulation by a similar factor (21x).  Since the cluster used far more vCPUs than are available in a single EC2 instance, Switchboard was the only way to parallelize the simulation to this degree.

Our final comparison entails using MT Verilator to produce a cycle-accurate ground truth against which we measure the accuracy of Switchboard simulations.  Since the largest MT Verilator simulation size was of an 8,192-core array, we scaled down the Switchboard simulation to match for this experiment.  The scaled-down setup consisted of a 4\texttimes 2 array of ECS tasks, each running a 4\texttimes 4 array of simulations, each of which was a single-netlist simulation of an 8\texttimes 8 CPU array.

The main control knob for accuracy in Switchboard is the maximum simulation rate.  As explained in Section~\ref{sec:perfsim}, performance measurements should approach ground truth if simulations are run at a controlled rate that is slow compared to real-world latencies between simulators.  Of course, it is only possible to slow down simulators with respect to their free-running rate, not speed them up.  Hence, to control the simulation speed, we set a maximum rate.

\begin{figure}
    \centering
    \includegraphics[width=1\linewidth]{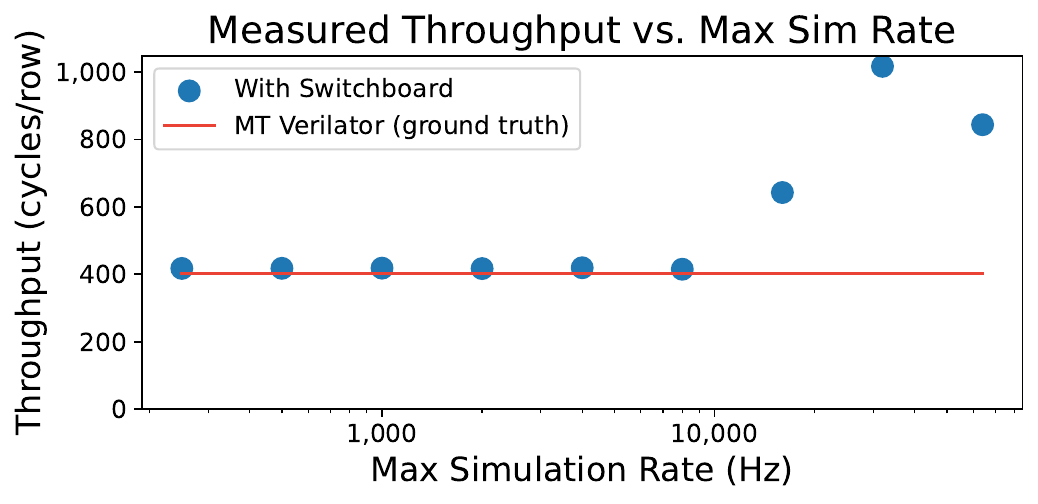}
    \caption{Effect of varying the Switchboard maximum simulation rate on measured throughput (accuracy indicator).}
    \label{fig:max-sim-rate}
\end{figure}

The performance measurement used in our accuracy study was the number of clock cycles taken to produce a single row of the matrix multiplication output.  We varied the max simulation rate over a wide range while taking this measurement, with the result shown in Figure~\ref{fig:max-sim-rate}.  As predicted, the performance measurement converged to a value close to ground truth as the max simulation rate was lowered, and for simulation rates of 8 kHz and lower, the accuracy with respect to ground truth was better than 5\%.

\section{Related Work}
\label{sec:related}

A broad range of techniques are used for simulating large hardware systems, including parallel RTL simulation, FPGA emulation, custom emulation hardware, application-level modeling, and modular simulation using one or more of these approaches.  In this section, we discuss related work using these techniques, which are geared towards digital designs, and briefly touch on the topic of mixed-signal simulation, which is related to Switchboard's support of SPICE models.

When an RTL implementation exists, parallel RTL simulation is appealing because it requires no special models and no modification of the design.  Most work in this area involves representing the RTL design as a computational graph, partitioning it into subgraphs, and assigning each subgraph to a thread.  Verilator has offered this type of parallelization since its 4.0 release~\cite{verilator4}, and since then others have sought to improve on its performance.  For example, RepCut~\cite{repcut} found that simulation performance could be boosted by 3x by duplicating small parts of the graph to eliminate data dependencies.  Parendi~\cite{parendi} targeted the manycore Graphcore IPU, enabling the efficient execution of a finely partitioned computational graph; the result was also a 3x speedup.

An alternative to RTL simulation is to map RTL to one or more FPGAs.  One of the open-source tools developed for this purpose is FireSim~\cite{FireSim}.  Like us, the authors implemented designs on AWS F1 instances, but their simulations were synchronized using an approach based on the Chandy-Misra-Bryant (CMB) algorithm~\cite{cmb-algorithm}.  FireSim was applied to the simulation of a 4,096-core data center, using 32 F1 instances and achieving a simulation rate of 3.4 MHz.  Another F1-based approach was SMAPPIC~\cite{smappic}, which achieved a higher simulation rate (100 MHz) in part by avoiding explicit synchronization, instead partitioning systems such that real-world communication latencies had appropriate relative sizes for the system being modeled.

RTL can also be mapped to purpose-built emulation hardware.  Commercial products in this space typically have a maximum capacity in the tens of billions of gates~\cite{veloce,zebu,palladium}, although cost and compile time can be high.  A recent approach, ASH~\cite{ash}, uses the Verilator frontend for fast compilation but targets a custom processor with speculative execution, achieving a 32x speedup over Verilator.

Another way to speed up simulations is to operate at the application binary level, using fast functional processor models in conjunction with cycle count performance models.  Work along these lines includes PriME~\cite{prime}, SimTRaX~\cite{simtrax}, and MuchiSim~\cite{muchisim}.  MuchiSim in particular demonstrated that this approach can scale to one million processing units.  A related approach is to simulate patterns of computation and memory access, as in MosaicSim~\cite{mosaicsim}, which used static analysis of LLVM IR and trace analysis to extract these patterns.

Modular simulation involves combining instances of one or more of the previous approaches to run a large simulation.  An example in this category is Metro-MPI~\cite{metro-mpi}, which split simulations into multiple processes using latency-insensitive interfaces as boundaries, as we have.  Like us, the authors observed that processes should represent large enough sections of the design to amortize overhead, calling these sections ``multi-tile granules'' (similar to our single-netlist simulations).  However, they targeted tightly-coupled compute platforms, using MPI to communicate between processes.  Processes were run in lockstep, synchronizing on each clock cycle.  The authors simulated OpenPiton+Ariane designs consisting of up to 1,024 cores (10B transistors) on a 48-core HPC system.  Compile time was 2h for the largest system; speedups of 35x in build time and 9.3x in run time were achieved with respect to Verilator.

Another modular simulation approach is SimBricks~\cite{simbricks}.  Like Switchboard, it connects simulation processes via SPSC shared-memory queues and supports heterogeneous simulation.  Each simulation process can be an instance of QEMU, gem5, Verilator, or one of several network simulators.  Boundaries between processes are drawn where there are PCIe or Ethernet interfaces in the system being modeled, and these interfaces are conveyed through queues.  As in FireSim, it is assumed that there is a non-zero latency associated with each interface, and processes synchronize using a CMB-like approach.  SimBricks was tested on systems of up to 1,000 simulated hosts, using up to 26 AWS instances.

Turning now to the problem of including analog models in hardware simulations, a commonly-used tool is Verilog-AMS~\cite{designers-guide}, which allows SPICE models to be instantiated in RTL, with digital and analog simulators running side-by-side.  However, simulation speed can be an issue, and there is no open-source implementation.  An alternative is to create real-number models (RNMs) in Verilog; work along these lines includes XMODEL~\cite{xmodel} and DaVE~\cite{dave}, which both avoid oversampling by operating on special analog waveform types.  Other work has proposed implementing RNMs alongside digital blocks on FPGAs~\cite{msdsl-paper} and on commercial emulators~\cite{Nothaft, zebu-ams}.

Switchboard synthesizes ideas from past work into an easy-to-use open-source framework, and is unique in several ways.  For one, it supports FPGA emulation and SPICE models, which other modular simulation frameworks do not.  In addition, Switchboard is able to scale to a degree not previously reported for RTL-based simulation, running a million-core RISC-V simulation (20B gates / 200B transistors) across hundreds of ECS tasks.  To our knowledge, this is the first use of ECS for distributed RTL simulation.  Finally, Switchboard introduces a new mechanism for approximate performance measurement without explicit synchronization, by controlling simulation rates.

\section{Conclusion}
\label{sec:conclusion}

Switchboard is based on a design methodology in which large hardware systems are composed of modular blocks that interact through latency-insensitive interfaces.  For designs that are structured this way, we propose prebuilding a model for each block, where a model can be an RTL simulation, RTL implemented on an FPGA, or a software model.  At runtime, one or more instances of each model are started, and they are connected through shared-memory queues, allowing systems of arbitrary size to be constructed without additional build time.  Due to the partitioning of the design at latency-insensitive interfaces, there is no need for explicit synchronization for functionally-correct simulations.  Approximate performance measurement is possible by slowing down simulations as needed such that the ratios of clock rates in wall time match the ratios of clock rates in simulation.

We demonstrated that there are several merits to this modular simulation approach.  For one, it allows systems to be modeled in a flexible manner, taking advantage of whatever models are available for the constituent blocks.  Due to the support of FPGA emulation, the use of shared-memory queues, and the lack of explicit synchronization, Switchboard-based modeling can be very fast, as demonstrated by the high-level interactive tasks supported by the web app.  Our approach is also highly scalable using standard cloud compute resources, as demonstrated by a simulation of one million RISC-V cores on thousands of cloud vCPUs spread across hundreds of ECS tasks.  Unlike approaches targeting a single tightly-coupled compute system, there is significant room to grow from here, and we believe this will be important as hardware designs themselves continue to scale.

\section*{Acknowledgment}
\label{sec:acknowledgement}

We thank everyone who contributed to the Switchboard open-source project and its applications, including Ben Sheffer, Ali Zaidy, Peter Gadfort, Peter Grossman, Wenting Zhang, J{\"u}rgen Leschner, Will Ransohoff, Amir Volk, and James Ross.

\bibliographystyle{IEEEtran}
\bibliography{IEEEabrv,references}


\begin{IEEEbiography}[{\includegraphics[width=1in,height=1.25in,clip,keepaspectratio]{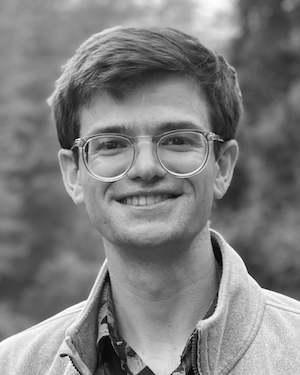}}]{Steven Herbst} is Director, Software Solutions at Zero ASIC, where he works on simulation technology for chiplet-based designs.  Steven completed a Ph.D. in Prof. Mark Horowitz's group at Stanford University (2021), where he worked on FPGA emulation of mixed-signal chip designs.  Prior to the Ph.D. program, he spent five years in industry as an analog designer for PCBs and ICs (2011-2016).  Steven holds S.B. and M.Eng. degrees in electrical engineering from MIT (2010, 2011).
\end{IEEEbiography}

\begin{IEEEbiography}[{\includegraphics[width=1in,height=1.25in,clip,keepaspectratio]{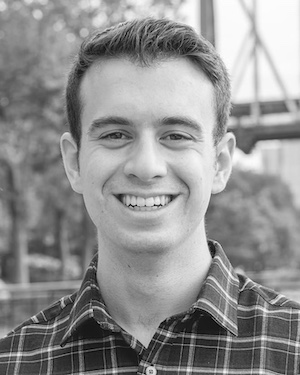}}]{Noah Moroze} is currently a software engineer at zeroRISC. He was previously at Zero ASIC (2021-2023), where he worked on a variety of projects including an emulation platform for chiplet-based systems and EDA build infrastructure. Noah completed his M.Eng. (2021) in MIT's Parallel and Distributed Operating Systems group, where he researched formal verification of security properties for embedded systems. Noah holds an S.B. in electrical engineering and computer science from MIT (2020). 
\end{IEEEbiography}

%\newpage

\begin{IEEEbiography}[{\includegraphics[width=1in,height=1.25in,clip,keepaspectratio]{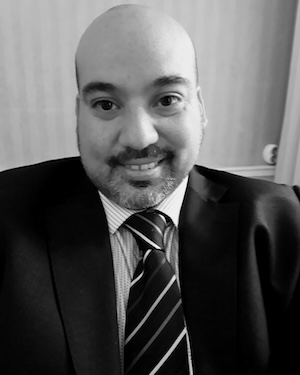}}]{Edgar Iglesias} is a Principal Software Engineer at AMD. He has over 25 years of experience working primarily in the areas of Embedded Systems, Virtual Platforms and Co-simulation. Over the years, he has contributed to numerous Open-Source projects (such as QEMU) and has previously given talks at the Embedded Linux Conference and the KVM Forum.
\end{IEEEbiography}

\begin{IEEEbiography}[{\includegraphics[width=1in,height=1.25in,clip,keepaspectratio]{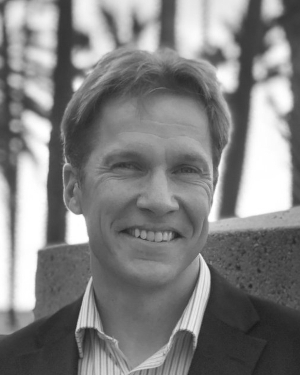}}]{Andreas Olofsson} is the founder and CEO at Zero ASIC, where he works on a variety of chiplet related design technologies. From 2017-2020, Andreas was a Program Manager at DARPA, where he managed national research programs in design automation, open source hardware, and heterogeneous integration. From 1997 to 2016, Andreas worked in commercial IC development,  participating in all aspects of the product life cycle, from architecture to post silicon automated testing. Andreas completed his B.A. in Physics (1996), B.Sc. in Electrical Engineering (1996), and M.Sc. in Electrical Engineering (1997) engineering from the University of Pennsylvania.
\end{IEEEbiography}


%\vfill


\end{document}